# Optical properties of ZnCoO films and nanopowders

**E. Wolska[1,2], M. Łukasiewicz[1], J.D. Fidelus[2], W. Łojkowski[2], E. Guziewicz[1], M. Godlewski[1,3]**

*[1]Institute of Physics, Polish Acad. of Sciences, Al. Lotników 32/46, 02-668 Warsaw, Poland*
*[2]Institute of High Pressure Physics, Polish Acad. of Sciences, "Unipress", Sokołowska 29/37, 01-142 Warsaw, Poland*
*[3]Dept. of Mathematics and Natural Sciences College of Science, Cardinal S. Wyszyński University, Dewajtis 5, 01-815 Warsaw, Poland*

## Abstract

ZnCoO is one of the most studied and promising semiconductor materials for spintronics applications. In this work we discuss optical and electrical properties of ZnCoO films and nanoparticles grown at low temperature by either Atomic Layer Deposition or by a microwave driven hydrothermal method. We report that doping with Cobalt quenches a visible photoluminescence (PL) of ZnO. We could observe a visible PL of ZnO only for samples with very low Co fractions (up to 1%). Mechanisms of PL quenching in ZnCoO are discussed. We also found that ZnO films remained n-type conductive after doping with Co, indicating that a high electron concentration and Cobalt 2+ charge state can coexist.

## 1. Introduction

ZnCoO is one of the most studied and promising semiconductor materials for spintronics applications. Two main reasons for this are: (a) $Co^{2+}$ has nearly an identical ionic radius as $Zn^{2+}$, and thus hopefully can be introduced to ZnO more uniformly and at a higher concentration [1], and (b) theoretical calculations predict ferromagnetic state in heavily *n*-type doped ZnCoO [2-3], whereas in the case of ZnMnO room temperature ferromagnetism was predicted in *p*-type samples [4].

The latter favors ZnCoO since we can control level of *n*-type doping of ZnO, but *p*-type doping is still difficult, and is not stable in time, limiting chances for a room temperature ferromagnetism of ZnMnO. Moreover, in ZnO Mn changes its charge state from 2+ to 3+ [5], making high *p*-type doping still more difficult. This is why we checked in this work possible charge states of Co in ZnO and also if films doped with Co remain n-type conductive, i.e., if a high electron concentration and $Co^{2+}$ charge state can coexist.

## 2. Experimental and results

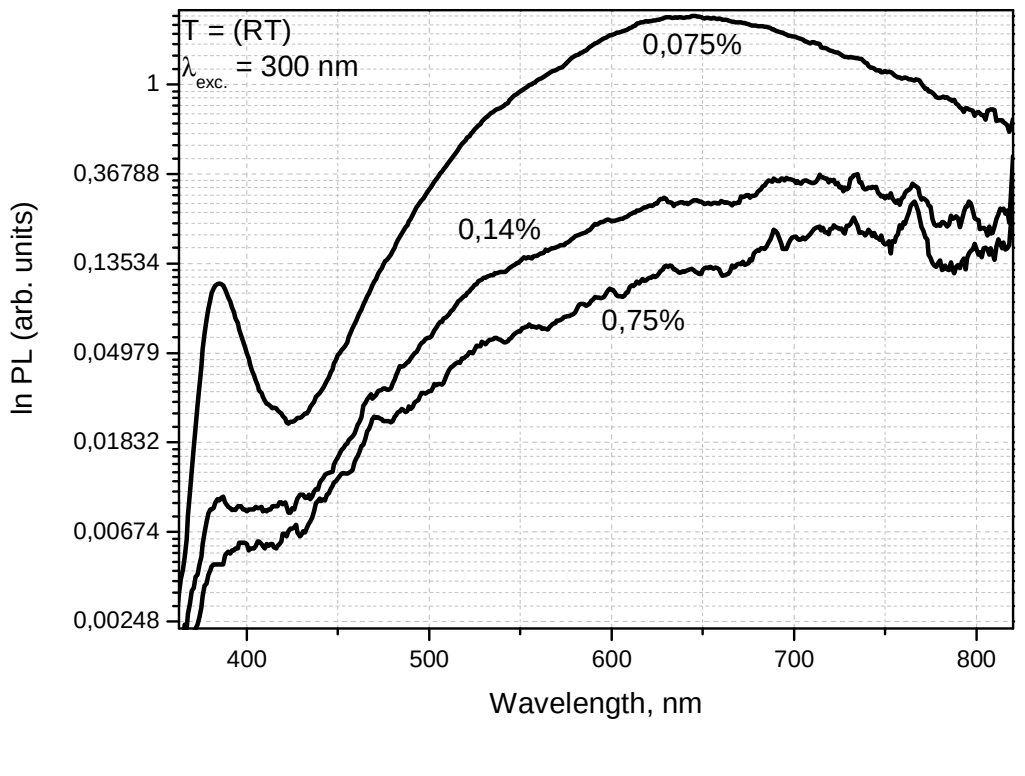

**Fig.1**: Visible PL of ZnCoO nanoparticles.

Three types of ZnCoO samples were studied – ZnCoO films grown by Atomic Layer Deposition (ALD), ZnCoO nanoparticles, and thin films obtained by spin coating of these nanoparticles, as described elsewhere ([1] and by M. Łukasiewicz, this conference). The samples contained low Co fractions of 0.1, 0.2 and 0.4 % for films grown by the ALD and 0.0075, 0.14 and 0.75 mol % for nanoparticles.

Optical properties of the samples were measured at room temperature.

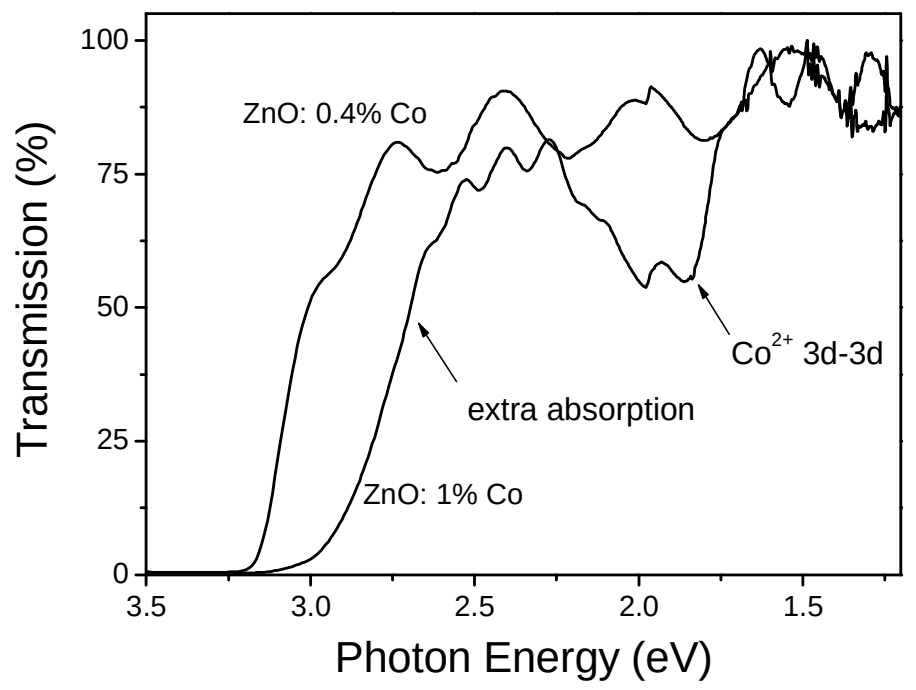

**Fig.2**: Transmission spectra of ZnCoO samples grown by the ALD.

In Fig.1 we show photoluminescence (PL) of ZnCoO nanoparticles for which quenching of both band edge and deep defect-related PL bands was observed. For films grown by the ALD only the band edge PL was observed, also effectively quenched in samples containing Cobalt.

We then measured transmission spectra, which are shown in Fig.2 for ZnCoO films grown by the ALD, and their electrical properties to evaluate effect of Co doping on free electron concentration. Free electron concentration (at room temperature) up to $10^{19}$ $cm^{-3}$ was found.

## 3. Discussion

Transition metal ions, such as Fe, Ni, Co, are effective killer centers quenching a visible PL of ZnS (see [6] and references given there). Present study performed for ZnCoO samples prepared on three different ways demonstrate (see Fig.1) that the similar situation occurs also in ZnO. A rapid quenching of a visible PL is observed for ZnCoO with Co fractions below 1 mol%, i.e., this process is more efficient that the one reported by us for ZnMnO films [5,7].

Mechanisms of a PL quenching by TM ions were reviewed by us is the reference [6]. Two the most efficient types of nonradiative processes are: (a) the so-called bypassing process [8], and (b) Auger-type processes, as described in the reference [6].

In the bypassing process free carriers are sequentially trapped by TM-related center and thus do not participate in processes resulted in a visible PL. Large efficiency of this process reflects large capture cross sections of free carriers by TM ions, which is the consequence of a strong electron - phonon coupling.

The data shown in Fig.1 clearly demonstrate the rapid quenching of ZnO visible PL upon Co doping. The similar effect was observed for three types of ZnCoO films studied in the present work, as is shown in the inset of Fig.1.

The bypassing process requires change of a charge state by a given TM ion. This commonly results is the appearance of a broad absorption band of photoionization character, as observed by us for e.g. Fe and Cr ions in ZnS [9,10]. In Fig.2 we compared transmission spectra measured for ZnCoO films with 0.4% and 1% Co fractions. In addition to the interference pattern a distinct extra absorption is resolved, being due to the $Co^{2+}$ intra - shell $^4A_2(^4F) \rightarrow {}^2A_1(^2G)$, $^4A_2(^4F)\rightarrow{}^2A_1(^4P)$ and $^4A_2(^4F) \rightarrow {}^2A_1(^2G)$ transitions, also reported by Jin et al. [11], Dinia et al. [12], Liu et al. [13], and Ramachandren et al. [14].

The second effect observed in Fig.2 is a red shift of the fundamental band gap absorption. We further assume that this shift is due to an overlap of the band-to-band and a photoionization transition. Cobalt-relation of this extra band was demonstrated by Jin et al. [11]. We propose that the bypassing process, responsible for efficient quenching of a ZnO visible PL, is due to a high efficiency of this photoionization transition.

Origin of the assumed photoionization is not clear. ESR investigations of other wide band gap II-VI compounds indicate two possibilities - either 2+ → 1+ or 2+ → 3+ Co recharging. ESR signal of $Co^{1+}$ ($3d^8$) has been observed in MgO, CaO and hexagonal CdS only [15]. In ZnO $Co^{2+}$ ESR signal was reported till now [16].

Concept of the universal reference level is usually used to predict possible charge states of various TM ions in different compounds [17]. Even though our previous studies of TM doped II -VI compounds indicate that the accuracy of this model is rather limited, this model suggests the possibility of 2+→3+ Co recharging in ZnO. The energy of the assumed 2+→3+ transition is estimated to be larger than 2.8 eV (thermal ionization energy), which is supported by observation of $Co^{2+}$ intra-shell transitions up to 2.2 eV.

The above assumption is in line with observation of relatively high free electron concentration in ZnCoO. In the case of $Co^{2+}\rightarrow Co^{1+}$ recharging doping with Co would compensate n-type conductivity and result in resistive samples.

Our previous investigated indicated that another possible explanation of PL quenching (Auger processes) is less likely [6].

## Acknowledgements:

This work was supported by FunDMS ERC Advanced Grant Research and (E. Wolska, J.D. Fidelus) grant no. N N508 0851 33 of Ministry of Science and Higher Education (Poland) donated for the years 2007-2009.